\documentclass[intlimits,twoside,a4paper]{article}
\usepackage{array}
\usepackage{amsmath,amssymb}
\usepackage{graphicx}
\usepackage[eqsecnum]{cmpj2}
\newcommand{\field}[1]{\mathbb{#1}}
\newcommand{\R}{\field{R}}
\newcommand{\Z}{\field{Z}}
\articletype{Regular article}
\title[$4D$ scalar field theory on a lattice] {Critical line of the $\Phi^4$ scalar field theory
                                   on a $4D$ cubic lattice in the local potential approximation.}
\author{Jean-Michel Caillol  \refaddr{label1,label2}}
\addresses{\addr{label1} Univ. Paris-Sud, Laboratoire LPT, UMR 8627, 
                         Orsay, F-91405, France
	   \addr{label2} CNRS, Orsay, F-91405, France		 
}
\begin{document}

\maketitle

\begin{abstract}
We establish the critical line of the  one-component  $\Phi^4$ (or Landau-Ginzburg) model 
on a simple  four dimensional cubic lattice.
Our study is performed in the framework
of the non-perturbative renormalization group in the local potential approximation with 
a soft infra-red regulator.  The transition is found to be of second order even
in the Gaussian limit where first order would be expected according to some recent
theoretical predictions.
\pacs 02.30.Jr;02.60.Lj;05.10.Cc;05.50.+q;64.60.De 
\end{abstract}
\section{\label{sec:intro}Introduction}
It is a real pleasure and a great honor for the author to contribute, with this paper,
to the festschrift dedicated to Professor Myroslav Holovko at the occasion of his
$70th$ birthday. Myroslav is an expert of the collective variables (CV) method introduced
by the Ukrainian school in the framework of which Wilson's ideas on the renormalization
group (RG)~\cite{Wilson} can be implemented with great effect~\cite{Yukhnovskii}. 
Here we expose recent post-Wilsonian advances on the RG in the framework of statistical field theory. Obviously, many of the ideas
exposed here could easily be transposed to the CV ``world'' by the readers of 
references~\cite{Caillol_CV_I,Caillol_CV_II} where the links between the CV method
and standard statistical field theory  are established.

These last past years, Wilson's approach  to the RG~\cite{Wilson,Wegner} has been the subject of a revival 
in both statistical physics and quantum field theory. Since the seminal work of Wilson,
two main formulations of 
the non-perturbative renormalization group (NPRG) 
have been developed in parallel.
Very similar to  the works of the Ukrainian school on the  of CV formalism 
we have the approaches initiated independently and in parallel  by Wetterich \textit{et al.} 
\cite{Wetterich1,Wetterich2,Delamotte,Ellwanger} on the one hand and Parola \textit{et al.} 
in the other hand~\cite{Parola1,Reatto,Caillol_III}. 
In this corpus of works one is interested to establish and to solve the 
flow equations of the Gibb's free energy  by means of non-perturbative methods.
In an alternative formulation, Polchinski and his followers consider  rather  the flow of the Wilsonian 
action~\cite{Polchinski,Bervillier}, instead of  that of the free energy,  which makes the method more 
abstract and less predictive than that of Wetterich, although  more in accord with Wilson's ideas.
The link between these two formulations can however be  established, see  for instance references~\cite{Morris,Caillol_0}.
Other non-perturbative methods based either on the CV or Monte Carlo  methods are also the subject of active
studies and are discussed, for instance,  in reference~\cite{Pylyuk} and references quoted herein.

The NPRG has proved its ability to describe \textbf{both universal and non universal} quantities 
for various models of statistical and condensed matter physics \textbf{near or even far} from criticality. 
Recently it has been extended to  models defined on a lattice~\cite{Dupuis-S}.
Successful applications  to the three-dimensional (3D) Ising, XY, Heisenberg models~\cite{Dupuis-M}
and $\Phi^4$ model~\cite{Caillol_II} are noteworthy. Here we extend the study of reference~\cite{Caillol_II} on the 
$\Phi^4$ model in three dimensions of space to the case $D=4$; it was made possible by the recent
publication by Loh of a novel numerical method to compute the lattice Green's functions~\cite{Loh}. The  $D=4$ version
of the $\Phi^4$ model on a lattice describes the field of a Higgs boson 
on the lattice in interaction with itself~\cite{Montvay}; our conclusions concerning the type of transition 
it undergoes are thus of theoretical importance.

As in our former study of the $D=3$ version of the model,  
we work in the framework of the local potential approximation  \cite{Wetterich1,Wetterich2,Delamotte,Caillol_II}
but here we consider  only the case of the Litim-Machado-Dupuis infra-red cut-off introduced in refs~\cite{Dupuis-M,Litim}
This regulator  has been shown to give much more better results than other sharp regulators in~\cite{Caillol_II}. 
As in  references~\cite{Parola1,Reatto,Bonanno,Caillol_II,Caillol_I} the flow equations are 
numerically integrated out for the so-called threshold functions~\cite{Wetterich2}
rather than for the potential. 
The resulting flow equations belong to the class of quasi-linear parabolic
partial differential equations (PDE) for which  several efficient and unconditionally convergent
numerical algorithms have been developed 
by mathematicians~\cite{Ames}.   As in references~\cite{Parola1,Reatto,Bonanno,Caillol_I,Caillol_II}
we made use of an algorithm proposed
by Douglas-Jones~\cite{Ames,Douglas}  to solve our NPRG flow equations, both \textbf{above}
and \textbf{below} the critical temperature; this yields an easy and precise determination of the critical point.
The critical line of the model is obtained for a large range of  parameters; unfortunately, and  contrary to
the case $D=3$~\cite{Caillol_II,Hasenbusch}, we were unable to
find available Monte Carlo simulations to which  to compare our data. 
We stress that, in the wide range of parameters considered in our study (see table~\ref{Tab}),  
we exclude the occurrence of a first order transition. This conclusion seems in agreement with a general analysis
of the criticality of the model made in reference~\cite{Hara_I,Hara_II}.

Our paper is organized a follows :
In section~\ref{secI} we review briefly the basic definitions and results 
concerning the statistical mechanics of  scalar fields on a lattice. Section~\ref{NPRG}
is devoted to theoretical and technical aspects of the NPRG on the lattice. We then present
our numerical experiments and discussed the results in section~\ref{Num}. We conclude
in section~\ref{Conclusion}
\section{\label{secI}Prolegomena}
\subsection{\label{model}Model}
Let us consider  some arbitrary  field theory defined on a $4D$ hyper-cubic
lattice
\begin{equation}
\label{def}
\Lambda = a \Z^{4}= \{ \mathbf{r} |\mathbf{r}_{\mu}/a \in \Z; \mu =1,\ldots,4 \}
\end{equation}
where $a$ is the lattice constant. The real, scalar field $\varphi_{\mathbf{r}}$ is defined on 
each point of the lattice. It is convenient to start with a finite hyper-cubic subset of points
 $\{ \mathbf{r} \} \subset\Lambda$ and to assume periodic boundary conditions (PBC) for the $\varphi_{\mathbf{r}}$
 before taking the infinite volume
limit, although no difficulties are expected to arise from this operation.

In  the case of short-range interactions between the fields, the action of the theory
can quite generally be written as \cite{Montvay}
\begin{equation}
\label{S1}
 \mathcal{S} \left[ \varphi \right] = \frac{1}{N a^4}  
 \sum_{\left\lbrace \mathbf{q} \in \mathcal{B}\right\rbrace }
 \widetilde{\varphi}_{-\mathbf{q}} 
\epsilon_0 \left(  \mathbf{q}   \right)  \widetilde{\varphi}_{\mathbf{q}}        + a^4
\sum_{\left\lbrace  \mathbf{r} \right\rbrace} U(\varphi_{\mathbf{r}} ) \; ,
\end{equation}
where $\left[ \varphi \right]$ is a shortcut notation for $\lbrace \varphi_{\mathbf{r}} \rbrace$ and 
 \begin{equation}
 \widetilde{\varphi}_{\mathbf{q}}=a^4 \sum_
{\left\lbrace  \mathbf{r} \right\rbrace}
e^{- i \mathbf{r}\mathbf{q}}  \varphi_{\mathbf{r}}
\end{equation}
is the Fourier transform of the field and the $N$ momenta $\left\lbrace \mathbf{q}\right\rbrace $ are
restricted to the  first Brillouin zone $\mathcal{B}= [-\pi/a, \pi/a ]^{\otimes4}$ of the reciprocal lattice.
The inverse transformation reads :  
\begin{equation}
\varphi_{\mathbf{r}}=  \frac{1}{Na^4}\sum_{\left\lbrace  \mathbf{q}
 \in \mathcal{B} \right\rbrace}
e^{ i \mathbf{r}\mathbf{q}}  \widetilde{\varphi}_{\mathbf{q}} \; .
\end{equation}
Note that, in the thermodynamic limit ($a$ fixed, $N \to \infty$),
$ \sum_{\left\lbrace \mathbf{q}\right\rbrace }  \rightarrow  (N a^4 )  \int_{ \mathbf{q}} $,  where
\mbox{$ \int_{ \mathbf{q} }  \equiv \int_{-\pi/a}^{\pi/a} \frac{d q_1}{2 \pi} \ldots \frac{d q_4}{2 \pi} $.}
In equation~(\ref{S1}) the spectrum 
$\epsilon_0 \left(  \mathbf{q}   \right)$ accounts for next-neighbor interactions. For a simple cubic (SC)
lattice it is equal to
\begin{equation}
 \epsilon_0 \left(  \mathbf{q}   \right) = (2 / a^2) \; \sum_{\mu=1}^{4}\left( 1-\cos\left(q_{\mu}a \right) \right) \; .  
 \end{equation}
Obviously one has 
$\epsilon_0 \left(  \mathbf{q}   \right) \sim  \mathbf{q}^2$
for $ \mathbf{q} \to 0$ and $\max_{ \mathbf{q}} \epsilon_0 \left(  \mathbf{q}\right) = \epsilon_0^{\max}=16/a^2$.
We will also define for convenience $k_{\mathrm{max}} \equiv 4/a$ by $\epsilon_0^{\max}= k_{\mathrm{max}}^2$.

Note that  in a system of units where the dimension of wave-vector $q_{\mu}$ is $ [q_{\mu}]=+1$, 
the dimension of the fields are $[\varphi_{\mathbf{r}}] = 1$ and $[\widetilde{\varphi}_{\mathbf{q}}] = -3$ 
so that the kinematic part of the action
$\mathcal{S} \left[ \varphi \right]$ is dimensionless.
Henceforth we shall only consider
the Landau-Ginzburg polynomial form $U(\varphi) = (r/2) \;  \varphi^2 + (g/4!) \; \varphi^4$. 
Since $[\varphi_{\mathbf{r}}] = 1$ and $[ a_{4}U(\varphi)]=0$ it follows that $[r]=2$ and $[g]=0$. Therefore,
in the thermodynamic limit,  the physics of the model  depends only upon the two dimensionless parameters
$\overline{ r}=  r a^2 $
and the dimensionless (only in $D=4$ ) $\overline{g} =g $.

Another way of writing the action \eqref{S1}, which is useful for numerical investigations, is \cite{Montvay,Hasenbusch}
\begin{equation}
 \label{S2}
 \mathcal{S} \left[ \psi \right] = \sum_{\left\lbrace \mathbf{n} \right\rbrace }
                                             \left[ -2 \kappa \sum_{\mu=1}^{4} \psi_{\mathbf{n}} \psi_{\mathbf{n}+\mathbf{e}_{\mu} }
                                                    + \psi_{\mathbf{n}}^2 
                                                    + \lambda \left(  \psi_{\mathbf{n}}^2 -1   \right) ^{2} -\lambda
\right]   \; ,
\end{equation}
where the $4$ unit vectors $\mathbf{e}_{\mu}$ constitute an orthogonal basis set for $\R^4$.
The field  $\psi$ and the parameters $(\kappa, \lambda)$ are all dimensionless and they are related to the bare field $\varphi$
and dimensionless parameters ($\overline{ r}$,  $g$) through the relations
\begin{subequations}
\label{toto}
\begin{align}
\label{toto_a}  \psi_{\mathbf{n}} &= \sqrt{\frac{1}{2 \kappa}}\; a \; \varphi_{\mathbf{r}}  \; \; \mathrm{ with } \; \; \mathbf{r} = a \mathbf{n} \; , \\
\overline{ r} &=  \dfrac{1 - 2 \lambda}{\kappa} -8    \; ,    \\ g &= \dfrac{6 \lambda}{\kappa^2} \; .
\end{align}
\end{subequations}
\subsection{\label{thermo}Thermodynamic and correlation functions}
The thermodynamic and structural properties of the model are coded in the partition function~\cite{Goldenfeld}
\begin{equation}
\label{Z}
 Z\left[  h \right] = \int \mathcal{D} \varphi \exp\left( -S\left[  \varphi \right] + \left(h \vert \varphi  \right)  \right)  \; ,
\end{equation}
where the dimensionless functional measure is given by
\begin{equation}
 \mathcal{D} \varphi = \prod_{\mathbf{n}}\; d\psi_{\mathbf{n}}  \; ,
\end{equation}
where  $\mathbf{r}= a \mathbf{n}$,  
the dimensionless $\psi_{\mathbf{n}}$ is defined at equation~\eqref{toto_a},  $h$ is an external lattice field,
and the dimensionless scalar product in~\eqref{Z} is defined as 
\begin{equation}
 \left(h \vert \varphi  \right)  = a^4 \sum_{\mathbf{r}} h_{\mathbf{r}} \varphi_{\mathbf{r}} \; .
\end{equation}

The order parameter is given by 
\begin{equation}
 \phi_{\mathbf{r}} =  \left\langle \varphi_{\mathbf{r}} \right\rangle = \dfrac{1}{a^4} 
 \frac{ \partial W\left[  h \right]}{\partial h_{\mathbf{r}}  } \; ,
\end{equation}
where the brackets $\langle \cdots\rangle$ denote statistical ensemble averages and the Helmholtz free energy 
$W\left[  h \right]= \ln Z\left[  h \right]$. Note that in the continuous limit, \textit{i.e.} $L=N a $ fixed, $ a \to 0$,
the partial derivatives tend to functional derivatives, \textit{i.e.} $ a^{-4}\partial  \cdots / \partial
h_{\mathbf{r}} \to \delta \cdots / \delta h(\mathbf{r})$.

It follows from  first principles that
$W\left[  h \right]$
is a convex function of the $N$
variables $\left\lbrace h_{\mathbf{r}}\right\rbrace$; it is also the
generator of the connected correlation functions $G^{(n)}(\mathbf{r}_1 \ldots \mathbf{r}_n) 
= a^{-4n} \partial^n  W\left[  h \right]/ \partial h_{\mathbf{r}_1}   \cdots \partial h_{\mathbf{r}_n}$,
where  $\partial \cdots / \partial h_{\mathbf{r}}$ denotes a partial derivative with respect
to one of the $N$  variables $h_{\mathbf{r}}$. 

The Legendre transform of $W\left[  h \right]$,
\textit{i. e. } the Gibbs free energy, will be provisionally 
denoted 
\begin{equation}
\label{G}
 \hat{\Gamma}\left[  \phi \right] = \left( h\vert \phi \right) - W\left[  h \right]  \; .
\end{equation}
$ \hat{\Gamma}\left[  \phi \right] $ is also -as a Legendre transform- a convex function of the $N$ conjugated
field variables  $\left\lbrace \phi_{\mathbf{r}} 
\right\rbrace $  It follows from equations~\eqref{Z} and \eqref{G} that the Gibbs potential is given
implicitly by the functional relation
\begin{equation}
 \label{impli_1}
\exp\left( -\hat{\Gamma} \left[ \phi \right]  \right) =
\int \mathcal{D}\varphi \; \exp \left( -\mathcal{S}\left[ \varphi \right] + (\varphi -
 \phi \; \vert \dfrac{\delta\hat{\Gamma}}{\delta \phi})
 \right) \; .
\end{equation}
where the abusive notation  $ \delta \cdots/ \delta \phi(\mathbf{r})\rightarrow
a^{-4} \partial \cdots / \partial \phi_{\mathbf{r}} $. has been used for a 
purpose of clarity.

The functional $ \hat{\Gamma}\left[  \phi \right]$  is 
 the generator of the so-called vertex functions $\hat{\Gamma} ^{(n)}(\mathbf{r}_1 \ldots \mathbf{r}_n) = a^{-4n} 
( \partial / \partial \phi_{\mathbf{r}_1} )   \ldots ( \partial / \partial \phi_{\mathbf{r}_n}) 
\hat{\Gamma}\left[  \phi \right] $. Finally, as well known~\cite{Goldenfeld},
the matrix  $\hat{\Gamma} ^{(2)}(\mathbf{r}_1 , \mathbf{r}_2)$ is the inverse of
matrix $G^{(2)}(\mathbf{r}_1 , \mathbf{r}_2)= \langle \varphi_{\mathbf{r}_1}\varphi_{\mathbf{r}_2}
\rangle - \langle \varphi_{\mathbf{r}_1}
\rangle\langle \varphi_{\mathbf{r}_2}
\rangle
$; \textit{i.e.} for $2$ arbitrary points
of the lattice $(\mathbf{x},\mathbf{y}) \in \Lambda $ one has 
\begin{equation}
\label{inv}
 a^4 \sum_{\mathbf{z} \in \Lambda} G^{(2)} (\mathbf{x},\mathbf{z} )
    \hat{\Gamma}^{(2)} (\mathbf{z},\mathbf{y} ) = 
\frac{1}{a^4} \delta_{\mathbf{x},\mathbf{y}} \; .
\end{equation}

\section{\label{NPRG} The State of art on Lattice NPRG}
\subsection{\label{NPRG_i}Lattice NPRG}
An elegant procedure to implement the lattice NPRG was given by Dupuis \textit{et al.} in references \cite{Dupuis-S, Dupuis-M};
it extends to the lattice the ideas of Wetterich \cite{Wetterich1,Wetterich2} for the continuum, \textit{i. e. } the
limit $a \to 0$  of the model;
it is very similar to the Reatto and Parola hierarchical reference theory of liquids \cite{Parola1,Reatto,Caillol_III}.
We add a quadratic term to the action~\eqref{S1} 
\begin{equation}
 \label{regu}
\Delta \mathcal{S}_{k}\left[   \varphi \right]  = \frac{1}{2} \frac{1}{N a^4} \sum_{\left\lbrace \mathbf{q}\right\rbrace }
                                     \varphi_{-\mathbf{q}} \widetilde{R}_k\left( \mathbf{q}\right)  \varphi_{\mathbf{q}}   \; .
\end{equation}
where $ \widetilde{R}_k\left( \mathbf{q}\right)$ is  positive-definite, has the dimension $[\widetilde{R}_k]=2$ and acts as a  $\mathbf{q}$ dependent mass term. 
The regulator $\widetilde{R}_k \left( \mathbf{q}\right) $ is chosen in such a way that it acts as an infra-red (IR) cut-off
which leaves the high-momentum modes unaffected
and gives a  mass to the low-energy ones.  Roughly $\widetilde{R}_k \left( \mathbf{q}\right) \sim 0 $ for $\vert \vert \mathbf{q} \vert \vert> k $
and $\widetilde{R}_k \left( \mathbf{q}\right) \sim Z_k k^2 $ for $\vert \vert \mathbf{q} \vert \vert <  k $.
The scale $k$ in momentum space varies from $\Lambda \sim a^{-1}$, some undefined microscopic scale of the  model yet to be defined precisely,
to $k=0$ the macroscopic scale. 
To each scale ``$k$'' corresponds a $k$-system defined by its microscopic action $ \mathcal{S}_{k}\left[   \varphi \right]=
\mathcal{S} \left[   \varphi \right] + \Delta \mathcal{S}_{k}\left[   \varphi \right] $.
We denote its partition function by $Z_k\left[ h \right] $, its Gibbs free energy by 
$ \hat{\Gamma}_k\left[  \phi \right]$, \textit{etc}. The generalization of equation~\eqref{impli_1} is then
\begin{equation}
 \label{Gimplicit}
\exp\left( -\Gamma_k\left[ \phi \right]  \right) =
\int \mathcal{D}\phi \; \exp \left( -\mathcal{S}\left[ \varphi \right] + (\varphi - \phi \; \vert \dfrac{\delta\Gamma_k[\phi]}{\delta \phi})
-\frac{1}{2} (\varphi - \phi\; \vert \widetilde{R}_k \vert \varphi - \phi\;)
 \right) \; ,
\end{equation}
where  the so-called average effective action
$\Gamma_k \left[\phi \right] $, which was introduced by Wetterich in the first stages of the NPRG, is 
defined as a modified Legendre transform of $W_k\left[ h \right] $ which includes the explicit subtraction of
$\Delta \mathcal{S}_{k}\left[   \phi \right]$ \cite{Wetterich1,Wetterich2}, \textit{i. e.}
\begin{equation}
  \Gamma_k \left[\phi \right] = \hat{\Gamma}_k\left[  \phi \right] -\Delta \mathcal{S}_{k}\left[   \phi \right] \; .
\end{equation}
Note that the functional $\Gamma_k \left[\phi \right] $ is not necessarily a convex functional of the classical field $\phi$
by contrast with $ \hat{\Gamma}\left[  \phi \right]$ which is the true Gibbs free energy of the  k-system. 

The choice of the regulator $\widetilde{R}_k\left( \mathbf{q}\right)$ would not affect exact results but matters as soon
as approximations are introduced.
We have retained the Litim-Dupuis-Machado (LMD)  regulator introduced by Dupuis and Machado~\cite{Dupuis-S, Dupuis-M}
for the lattice as an extension of Litim's regulator widely used for off-lattice field theories~\cite{Litim}. 
Sharp cut-off regulators often yield unphysical
behaviors, notably in the local potential approximation,
and should be avoided, see \textit{e. g.} \cite{Caillol_I,Caillol_II}.
The LMD regulator reads
\begin{equation}
 \label{Litim}
 \widetilde{R}_k\left( \mathbf{q}\right)= \left(
 \epsilon_k -\epsilon_0\left(\mathbf{q}\right)\right) \Theta\left( 
 \epsilon_k -\epsilon_0\left(\mathbf{q}\right)
 \right) \; ,
\end{equation}
where $\epsilon_k = k^2$ and $\Theta$ is the Heavyside's step function.
At scale ``$k$'', the effective spectrum of the $k-$model of action 
$\mathcal{S}_{k}\left[   \varphi \right]$ is clearly
\begin{equation}
\label{eps_eff}
 \epsilon_k^{\mathrm{eff.}} (\mathbf{q})=\epsilon_0\left(\mathbf{q}\right) +
 \left(
 \epsilon_k -\epsilon_0\left(\mathbf{q}\right)\right) \Theta\left(   
 \epsilon_k -\epsilon_0\left(\mathbf{q}\right)
 \right) \; .
\end{equation}
We note that for $\epsilon_0\left(\mathbf{q}\right) >\epsilon_k $ the regulator 
$\widetilde{R}_k\left( \mathbf{q}\right)$ vanishes in agreement with the fact that the high energy modes are of affected,
\textit{i. e.} one has $\epsilon_k^{\mathrm{eff.}} (\mathbf{q})=\epsilon_0\left(\mathbf{q}\right)$.
Conversely, for $\epsilon_0\left(\mathbf{q}\right) <\epsilon_k $, a constant massive contribution
is associated to the low-energy
modes, with a tendency to a freezing of their fluctuations, \textit{i. e.} one has
$\epsilon_k^{\mathrm{eff.}} (\mathbf{q})=\epsilon_k$.

 It is easy to show the the average effective action  satisfies the exact flow equation \cite{Wetterich1,Wetterich2,Delamotte,Dupuis-S,Dupuis-M}
\begin{equation}
 \label{flow}
\partial_k \; \Gamma_k\left[ \phi \right] = \frac{1}{2}  \; \sum_{\mathbf{q} \in \mathcal{B}} \partial_k \widetilde{R}_k\left( \mathbf{q}\right) 
                         \widetilde{G} ^{(2)}_{k}  (\mathbf{q}, -\mathbf{q} ) \; ,
\end{equation}
where   $ \widetilde{G} ^{(2)}_{k}$ is the Fourier transform of connected pair correlation function of the $k-$system defined as
\begin{equation}
 \widetilde{G} ^{(2)}_{k}( \mathbf{p},  \mathbf{q} ) = a^8 \; \sum_{\mathbf{x},\mathbf{y} \in \Lambda  }
                                                                                \exp\left( \ri  \mathbf{p} \cdotp  \mathbf{x} 
                                                                                +  \ri  \mathbf{q} \cdotp  \mathbf{y}  \right)
                                                                            G ^{(2)}_{k}( \mathbf{x},  \mathbf{y} )  \; .
\end{equation}

For an homogeneous configuration of the field $\phi_{\mathbf{r}}= \phi$ we have, on the one hand, 
 $\Gamma_k\left[ \phi \right]= N a^4 U_k(\phi) $ where
the potential $U_k(\phi)$ is a simple function of the field $\phi$ and, on the other hand, the conservation of momentum 
at each vertex which implies, with the usual 
abusive notation,  $ \widetilde{G}_k^{(2)}(\mathbf{q},- \mathbf{q})= N a^4  \widetilde{G}_k^{(2)}(\mathbf{q})$; from these 
remarks it follows that :

\begin{subequations}
\label{flow_2}
\begin{align}
\partial_k U_k ( \phi )  &= \frac{1}{2}\frac{1}{N a^4}  \; \sum_{\mathbf{q}} \frac{\partial_k \widetilde{R}_k\left( 
\mathbf{q}\right) }{   \widetilde{\Gamma}_k^{(2)}(\mathbf{q})
                    + \widetilde{R}_k\left( \mathbf{q}\right)} \; , \\
&=  \frac{1}{2}  \int_{\mathbf{q} \in \mathcal{B}} \frac{\partial_k \widetilde{R}_k\left( \mathbf{q}\right) }{   
\widetilde{\Gamma}_k^{(2)}(\mathbf{q})
                    + \widetilde{R}_k\left( \mathbf{q}\right)} \; , \label{limit_TH}
\end{align}
\end{subequations}
where the second line~\eqref{limit_TH} is valid  in the thermodynamic limit ($a$ fixed, $N \to \infty$). Note that in order
to establish equation~\eqref{flow_2} we also took into account of equation~\eqref{inv} in Fourier space for the $k-$system,
\textit{i.e.}  $  \widetilde{G}_k^{(2)}(\mathbf{q}) = 1/[ \widetilde{\Gamma}_k^{(2)}(\mathbf{q})
                    + \widetilde{R}_k\left( \mathbf{q}\right)]$, for an homogeneous system.
The reader will agree  that equation.~\eqref{flow_2}, which is \textit{exact},  is an extremely 
complicated equation since the vertex function
 $ \widetilde{\Gamma}_k^{(2)}(\mathbf{q},-\mathbf{q})$,
 which is the Fourier transform of the second-order functional derivative
of $ \widetilde{\Gamma}\left[ \phi \right] $ with respect to the classical field $\phi$, depends functionally upon  $\phi$. 

The  implicit solution~\eqref{Gimplicit}  of~\eqref{flow_2} allows us to establish precisely  the initial conditions. The initial
value $k= \Lambda$ of the momentum scale $k$ of the flow  is chosen such that $\widetilde{R}_{\Lambda}(\mathbf{q}) \sim \infty$ 
for all values of $\mathbf{q}$ hence,
since $\exp( -1/2 \; (\chi \vert \widetilde{R}_{\Lambda} \vert \chi )) \propto \delta[\chi] $,  where $\delta[\chi]$ is the Dirac functional, 
it follows from~\eqref{Gimplicit} that
$\Gamma_{\Lambda}[ \phi] =\mathcal{S}[ \phi]$.
Physically it means that all fluctuations are frozen and the mean-field theory becomes exact. 
When the running momentum goes from $k=\Lambda$
to $k=0$ all the modes $\widetilde{\varphi}_{\mathbf{q}}$ are integrated out progressively and the effective 
average action evolves from its microscopic limit
$\Gamma_{\Lambda}[ \phi] = \mathcal{S}[ \phi]$ to its final macroscopic expression $\Gamma_{k=0}[ \phi] =\Gamma[ \phi]$. 
\subsection{\label{loc}Local models and the initial condition of the flow}
Some members of our family of $k$-systems are nice fellows. It follows from~\eqref{eps_eff} that,
for $\Lambda >k > k_{\mathrm{max}} $, or equivalently $\epsilon_k >\epsilon_0\left(\mathbf{q}\right)$
for all vectors $\mathbf{q}$ of the first Brillouin zone, 
we have  $\epsilon_k^{\mathrm{eff.}} \equiv \epsilon_k$ which means that the action $\mathcal{S}_{k}\left[   \varphi \right]$
of the $k$-system is local and reads 
$\mathcal{S}_{k}\left[   \varphi \right]=a^4 \sum_{\left\lbrace  \mathbf{r} \right\rbrace}
\left[ U(\varphi_{\mathbf{r}}) + (1/2) \; \epsilon_k \; \varphi_{\mathbf{r}}^2    \right] $.
Therefore, at scale ``k'', we have a theory of independent fields on a lattice, which is trivial. 

The partition function  $Z_k\left[ h \right]= \prod_{\mathbf{r}}  z_k(\overline{h}_{\mathbf{r}})$
is a product of one-site partition functions with
\begin{equation}
\label{z}
 z_k(\overline{h}) = \int_{-\infty}^{ +\infty} d\overline{\varphi}\; \exp \left( 
 -\overline{U}(\overline{\varphi})  - \frac{1}{2} \overline{\epsilon}_k \overline{\varphi}^2 + \overline{h}\overline{\varphi} \right)  \; ,
\end{equation}
where we have introduced the dimensionless variables $\overline{\varphi}= a \varphi$, $\overline{h}= a^3 h$, and 
$\overline{\epsilon}_k= a^2 \epsilon_k$. Note that $U (\varphi)=a^4 \; \overline{U}(\overline{\varphi})$. The Helmholtz free energy
and Wetterich effective action can be written as lattice sums
\begin{subequations}
\label{latt}
\begin{align}
 W_k[h] &= \sum_{\mathbf{r}}   \ln  z_k(\overline{h}_{\mathbf{r}})                     \\
 \Gamma_k[ \phi]       &=  \sum_{\mathbf{r}} \gamma_k(\overline{\phi}_{\mathbf{r}})   \; ,
\end{align}
\end{subequations}
where the convex functions $\ln  z_k(\overline{h})$ and $\gamma_k(\overline{\phi})$ are related by a Legendre 
transform $\gamma_k(\overline{\phi}) + \ln  z_k(\overline{h})=\overline{\phi}\;\overline{h}  $, with, for instance
$\overline{\phi} = d\ln  z_k(\overline{h})/d\;\overline{h} $.
In general, the quantities $\ln  z_k(\overline{h})$ and $\gamma_k(\overline{\phi})$ cannot be computed
analytically but can easily be evaluated numerically for any value of $\Lambda >k > k_{\mathrm{max}} $.

It is interesting to note that the implicit equation~\eqref{Gimplicit} now reads
\begin{equation}
\label{local_implicit}
 \exp\left(- \gamma_k\left(\overline{\phi} \right)  \right) = \int_{-\infty}^{ +\infty} d\overline{\varphi} \;
\exp\left(
-\overline{U}(\overline{\phi}) + (d \gamma_k(\overline{\phi})/ d\overline{\phi}) \;  (\overline{\varphi} - \overline{\phi}) - 
\frac{1}{2} \; \overline{\epsilon}_k \;  \overline{\varphi}^2  
 \right)  \; ,
\end{equation}
which leads us to two remarks. First, the choice $\Lambda = \infty$ implies $\gamma_\Lambda=U$ since we can replace 
the Gaussian $\exp(-(1/2) \; \overline{\epsilon}_{\Lambda} \; \overline{\varphi}^2)$
by a delta function $\delta(\varphi)$ in equation~\eqref{local_implicit}. Our initial condition for the flow
of $\Gamma_k[\phi]$ is now perfectly defined.

Our second remark is that one can derive from the equation~\eqref{local_implicit}, \textit{i.e.} from
his solution!, the flow equation in the range
$\Lambda >k > k_{\mathrm{max}}$. A short calculation reveals that
\begin{equation}
 k \partial_k\gamma_k\left(\overline{\phi} \right)= \frac{\overline{\epsilon}_k}{\overline{\epsilon}_k+
 \gamma_k^{''}\left(\overline{\phi}\right) } \; .
\end{equation}

Noting that, for a homogeneous system, $\gamma_k(\overline{\phi}) = a^4 U_k(\phi)$, where $U_k(\phi)$
is the local potential defined in previous section~\ref{NPRG_i} the flow equation for the local potential
reads
\begin{equation}
\label{zou}
  \partial_{t}U_k = - \frac{a^4 \epsilon_k}{\epsilon_k + U_k^{''}}  \; ,
\end{equation}
with $\partial_{t}= -k\partial_{k}$. Clearly, equation~\eqref{zou} can also be obtained directly from~\eqref{flow_2}
in the range $\Lambda >k > k_{\mathrm{max}}$.

We are now in position to explicate the initial conditions which can be used to solve the 
flow equation~\eqref{flow_2} for the local potential 
\begin{itemize}
 \item either $ \Lambda = \infty $ and $U_{\Lambda} = U$ (Mean field theory like initial conditions). 
       In this case the flow
       equation~\eqref{zou} must be solved numerically for $\Lambda >k > k_{\mathrm{max}}$.
       Note that $U_{\Lambda}(\phi)$ can be non-convex.
 \item or  $\Lambda= k_{\mathrm{max}}= 4 / a $ and  $ U_{\Lambda}(\phi)  \equiv a^{-4}
 \gamma_{\Lambda}(\overline{\phi}) $ \; . In this case $\gamma_{\Lambda}(\overline{\phi})$ must
 be evaluated numerically (Local field theory like initial conditions). 
 Note that $U_{\Lambda}(\phi)$  is necessarily convex.
\end{itemize}
In our numerical experiments  we retained the second term of the alternative.
\subsection{\label{local}The local potential approximation}
\subsubsection{\label{general}The general case}
A non-perturbative, but intuitive approximation  to solve the flow equation.~\eqref{flow_2} is to make an ansatz on the functional form of 
$\Gamma_k[\phi]$. In the  local potential approximation (LPA)
one neglects the renormalization of the spectrum and assume that \cite{Dupuis-S,Dupuis-M}
\begin{equation}
 \label{LPA}
\textrm{(LPA ansatz)} \; \; \Gamma_k\left[ \phi \right] =  \frac{1}{N a^4}  \sum_{\left\lbrace \mathbf{q}\right\rbrace } \phi_{-\mathbf{q}} 
                                           \epsilon_0  \left(  \mathbf{q}   \right)  \phi_{\mathbf{q}}        + a^4
\sum_{\left\lbrace  \mathbf{r} \right\rbrace} U_k(\phi_{\mathbf{r}} ) \; .
\end{equation}
For a uniform configuration of the classical field $\phi_{\mathbf{r}} = \phi$ and, in the thermodynamic limit, 
 the flow equation~\eqref{limit_TH} becomes : 
\begin{equation}
\label{flow_LPA}
\partial_k U_k ( \phi ) 
 =  \frac{1}{2}  \int_{\mathbf{q} \in \mathcal{B}} \frac{\partial_k \widetilde{R}_k\left( \mathbf{q}\right) }{  \epsilon_0(\mathbf{q})
                    + \widetilde{R}_k\left( \mathbf{q}\right) + U_{k}^{''}(\phi)} \; ,
\end{equation}
where $U_{k}^{''}(\phi)$ denotes the second-order derivation of $U_k(\phi)$ with respect to the order parameter $\phi$.
Equation~\eqref{flow_LPA} is a non-linear parabolic PDE. These are good news since mathematicians have 
worked hard to provide us with numerical methods for solving such equations.
The equation must be supplemented by  initial and boundary conditions which will be explicated in section~\ref{Num_I}
\subsubsection{\label{local_LMD}The LMD regulator}
With the LMD regulator~\eqref{Litim} the loop-integral  in the r.h.s. of equation~\eqref{flow_LPA} can be worked out
analytically 
which leaves us with a much simplified flow equation for the potential  
\begin{equation}
\label{flow_LMD_final}
 \partial_t U_k= - \mathcal{N}(\epsilon_k) \mathcal{L}(\omega_k) \; ,
\end{equation}
where the RG time "$t$" is defined by $k=\Lambda e^{-t}$, so that $\partial_t = -k \partial_k$,
$\omega_k (\phi) \equiv U_k^{''}(\phi) /\epsilon_k$ is a dimensionless renormalized inverse 
susceptibility,
\begin{equation}
\label{L}
 \mathcal{L}(x) = \frac{1}{1 + x} \; ,
\end{equation}
is the threshold function \cite{Wetterich2} which takes a very simple expression with the LMD
regulator and finally 
\begin{equation}
 \mathcal{N}(\epsilon)  = \int_{\mathbf{q}\in \mathcal{B}} \Theta (\epsilon - \epsilon_0(\mathbf{q}))
\end{equation}
denotes the (normalized) number of states (note that we set $a=1$ to simplify the algebra).
It proves convenient to introduce also the density of states
\begin{equation}
\label{D}
 \mathcal{D}(\epsilon)  = \int_{\mathbf{q} \in \mathcal{B}} \delta (\epsilon - \epsilon_0(\mathbf{q})) \; ,
\end{equation}
so that 
\begin{equation}
  \mathcal{N}(\epsilon)  = \int_{0}^{\epsilon} \; d \epsilon^{'} \;  \mathcal{D}(\epsilon^{'})  \; .
\end{equation}
\begin{figure}[t!]
\centering
\includegraphics[angle=0,scale=0.40]{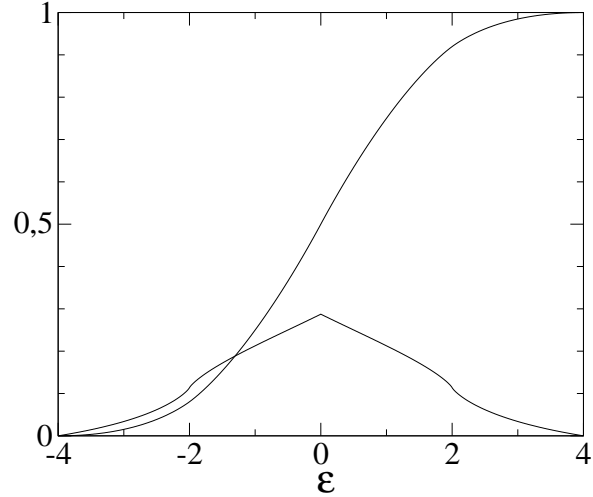}
\caption{
\label{D_N} Density and number of states, respectively $\mathcal{D}(\epsilon)$ (bottom) and 
$\mathcal{N}(\epsilon)$ (top), for the simple $D=4$ cubic lattice.
}
\end{figure}
The two functions  $\mathcal{D}(\epsilon)$ and $\mathcal{N}(\epsilon)$ are obviously related to the lattice Green function 
which, for a SC lattice,  reads \cite{JAP1,JAP2,Loh}
\begin{equation}
\label{Green}
 G(\tau) = \frac{1}{\pi^4} \int_0^{\pi}dq_1 \ldots \int_0^{\pi}dq_4  \frac{1}{\tau - \sum_{\mu = 1}^{\mu = 4} \cos(q_{\mu})}  \; .
\end{equation}
Note that we have, in the sense of distributions, for $\eta \to 0+$,  $1/(\tau + \ri \eta ) =
\mathcal{P}(1/\tau)+ \ri \pi \delta(\tau)$, where $\mathcal{P}$ is Cauchy principal part. 
With this remark, the comparison of equations~\eqref{D} and~\eqref{Green}
reveals at first sight that
\begin{equation}
 \mathcal{D}(\epsilon) = \frac{1}{2 a^2 } \frac{1}{\pi} \Im{G} (\tau) \; ,
\end{equation}
with $\tau = 4 - a^2/2\epsilon$. Note that the interval of the spectrum
$0 \leq \epsilon_k \leq \epsilon_0^{\mathrm{max}}$ corresponds
to the interval $-4 \leq \tau \leq 4$ for the auxiliary variable $\tau$.
Recently, in reference~\cite{Loh}, Loh has obtained a novel integral representation of
the Green's function of simple hyper cubic lattices. The resulting one-dimensional 
integral obtained for $G(\tau)$ involves non-oscillating, well behaved functions 
and it can thus be computed  precisely by means of a Gauss quadrature.
From the results of reference~\cite{Loh}
we obtained
\begin{itemize}
 \item For $0 \leq \epsilon \leq 2 $
 \begin{subequations}
 \label{zto}
 \begin{align}
  \mathcal{N}(\epsilon) = & \frac{1}{2} +  \int_0^{\infty}p_{02}(\epsilon,x) dx \\
  p_{02}(\epsilon,x)    = & \frac{1}{4\pi} I_e(x) K_e(x)^3 \frac{[3 \exp(-2 x) - 
                                                                  \exp((\epsilon-2)x)
                                                      -2 \exp(-(\epsilon+2)x) ]}{\epsilon} \; ,
 \end{align} 
 \end{subequations}
  where $I_e(x)=I_0(x) \exp(-x)$, $K_e(x)=K_0(x) \exp(x)$, $I_0(x)$ and  $K_0(x)$ being the 
 modified Bessel Functions of first and second class respectively.
 \item For $2 \leq \epsilon \leq 4 $
  \begin{subequations}
  \label{ztp}
 \begin{align}
  \mathcal{N}(\epsilon) = & 1 -  \int_0^{\infty}p_{24}(\epsilon,x) dx \\
  p_{24}(\epsilon,x)    = & \frac{I_e(x) K_e(x)}{4\pi} \; [I_e(x)^2 \frac{\exp((2 -\epsilon)x) -\exp(-2 x) }{\epsilon} \nonumber \\
                           &                           -K_e(x)^2 \frac{\exp(-(2+ \epsilon)x) -\exp(-6 x) }{\epsilon} ]\; ,
 \end{align} 
 \end{subequations}
 while, for negative values of $\epsilon$, one uses $\mathcal{N}(-|\epsilon|)= 1 - \mathcal{N}(|\epsilon|)$
 and one of the equations~\eqref{zto} or ~\eqref{ztp}.
\end{itemize}
The functions $\mathcal{N}(\epsilon)$ and $\mathcal{D}(\epsilon)$ were computed from the expressions~\eqref{zto}
and~\eqref{ztp} and are displayed in figure~\eqref{D_N}.
The Bessel functions involved in equations~\eqref{zto}
and~\eqref{ztp} were evaluated with the double-precision FORTRAN codes 
$\mathrm{i0}$ and $\mathrm{k0}$ of the specfun library
of the Netlib distribution~\cite{Netlib} while we made use of the code $\mathrm{DQAGIE}$ of the quadpack library,
of the same distribution, for the numerical integrations.

\subsection{\label{Structure} Various limits}
We first note that, for $\infty > k > k_{\mathrm{max}}$ one has the trivial identity 
$\mathcal{N}(\epsilon_k)=a^{-4}$.
Therefore the LMD flow equation~\eqref{flow_LMD_final} is identical to the
exact NPRG equation~\eqref{zou} for the local potential. LMD approximation is thus exact
for local theories~\cite{Caillol_II}.

Secondly we consider the scaling limit $k \to 0 $. We have
\begin{equation}
 \mathcal{N}(\epsilon)  = \int_{\mathbf{q}\in \mathcal{B}} \Theta (\epsilon - \epsilon_0(\mathbf{q}))
\sim  \int_{\mathbf{q}\in \mathcal{B}} \Theta (k^2 - \mathbf{q}^2))
                        \sim v_4 k^4 \;
\end{equation}
where $v_4=1/(32 \pi^2)$ is a geometrical factor,
then,  the flow equation~\eqref{flow_LMD_final} reduces to
\begin{equation}
\label{flow_LMD_asympt}
 \partial_t U_k = - v_4  k^4  \mathcal{L}(\omega_k)  \; \; (k \to 0) \; ,
\end{equation}
which is of course the LPA flow equation for the continuous (off-lattice)  theory
with Litim regulator~\cite{Ni1,Ni2,Ni3,Caillol_I}.
In the scaling limit, the lattice and off-lattice versions of  the $\Phi^4$ model share the same 
fixed-points and critical exponents, if any.

Let us discuss briefly the Gaussian fixed points solutions of equation~\eqref{flow_LMD_asympt}.
A general discussion, \text{i.e.} for arbitrary dimension $D$ and regulator $\mathcal{L}$, can be found in
reference~\cite{Caillol_I} while the case of a sharp cut-off was discussed for the first time
in the inspiring paper of Hasenfratz-Hazenfratz~\cite{Hasen}.

Fixed point solutions make sense only for an equation involving 
but dimensionless functions and variables and
emerge in general the limit $k \to 0$.
We introduce the dimensionless field $x=k^{-1}\phi$ and potential $u_k(x)=k^{-4}U_k(\phi)$. The adimensioned
flow equation can thus be written
\begin{equation}
\label{adim}
 \partial_t u_k = 4 u_k - x u_k^{'} - \frac{v_4}{1+u_k^{''}} \; ,
\end{equation}
with $u_k^{'} \equiv du_k/dx$. A fixed point $u^{\star}(x)$ satisfies
$\partial_t u^{\star}(x) =0$ for all $x$. $u^{\star \, ''}(x)=0$ is obviously
a special solution. By integration it gives $u^{\star \, '}(x)=0$ ($\mathbb{Z} \, 2$ symmetry)
and $u^{\star}(x)=v_4/4$, this is the Gaussian fixed point.
In order to study the stability of the fixed point
we linearize~\eqref{adim}. Let us define
\begin{equation}
 u_k(x)=u^{\star}(x)+ h_k(x) 
\end{equation}
and expand equation~\eqref{adim} in powers of $h$, it yields
\begin{subequations}
\begin{align}
 \partial_t h &= Dh -v_4 h^{'' \, 2}  \label{nkm} \\
 Dh   &=4 h -x h^{'} + v_4h^{''}  \; .
\end{align}
\end{subequations}
Let us start the analysis with the linearized RG equation
\begin{equation}
\label{linear}
 \partial_t h = Dh \; .
\end{equation}
We search a solution under the form $h(x,t)=\exp(\lambda t) H(y=\beta x)$
which yields the eigenvalue problem $(D -\lambda) H = 0$ which can be rewritten as
Hermite equation :
\begin{equation}\label{H}
 H^{''}(y) - 2y \, H^{'}(y) +2n \, H(y) =0 \; ,
\end{equation}
with $4- \lambda =n$. Hermite's equation~\eqref{H} admits in general
solutions without definite parity (Weber's functions). Only if $n$ is a
positive integer do
the solutions $H_n(y)$ have the same parity as $n$. Such solutions
are polynomials, namely the Hermite's polynomials~\cite{Morse}. Imposing $\Z 2$ symmetry
therefore leads to a discretization of the spectrum $4 - \lambda_p=2 p$, $p$
positive integer. The general linearized solution of~\eqref{linear} is then
\begin{subequations}
 \begin{align}
   h(x,t) &=\sum_{p=0}^{\infty} c_p \exp(\lambda_p t) \; H_{2p}(x/\sqrt{2 v_4}) \; ,\\ 
          &=\sum_{p=0}^{\infty} \widehat{c}_p \exp(\lambda_p t)\; \chi_p(x) \; ,
 \end{align}
\end{subequations}
where $\chi_p(x)$ is a convenient redefinition of Hermite's polynomial $H_{2p}$ such that
its coefficient of degree $2p$ is one. We have $\chi_0(x)=1$, $\chi_1(x)=x^2-v_4/2$,
$\chi_2(x)=x^4-6 v_4 x^2 + 3 v_4^2$, etc

Clearly for $p=0$ we have a trivial constant solution. $p=1$ corresponds to $\lambda_1=2$
thus $\chi_1(x)$ is a relevant field. 
The case $p=2$ corresponds to $\lambda_2=0$ and 
$\chi_2(x)$ is a marginal field. For all $p \geq 3$ the eigenvalue $\lambda_p <0$ (for
instance $\lambda_3=-2$) correspond to irrelevant solutions $\chi_p(x)$.
The stability of the marginal field $\chi_2(x)$ can be obtained by finding a 
solution of equation~\eqref{nkm} equal to $\chi_2$ at the dominant order. An analysis
similar to that of reference~\cite{Hasen} reveal that $\chi_2$ is in fact irrelevant
beyond the linear approximation. The picture of the scaling fields $\chi_p(x)$ in $D=4$
is thus consistent with a critical point~\cite{Goldenfeld}. The usual analysis~\cite{Goldenfeld}
then yields for the critical exponent $\nu$ the classical value $\nu=1/\lambda_1=0.5$.
Since Fisher's exponent $\eta=0$ in the LPA all other (classical) exponents are deduced from
scaling relations.

It is generally admitted, and was confirmed by the recent numerical studies of Codello~\cite{Codello}
that there is no other fixed point than the Gaussian fixed point in $D=4$. 
We have just shown that the LPA/LMD theory, albeit approximate, supports the existence 
of this fixed point.
\section{\label{Num}Numerical experiments}
\subsection{\label{Num_I}A change of variables}
\begin{figure}[!]
\centering
\includegraphics[angle=0,scale=0.40]{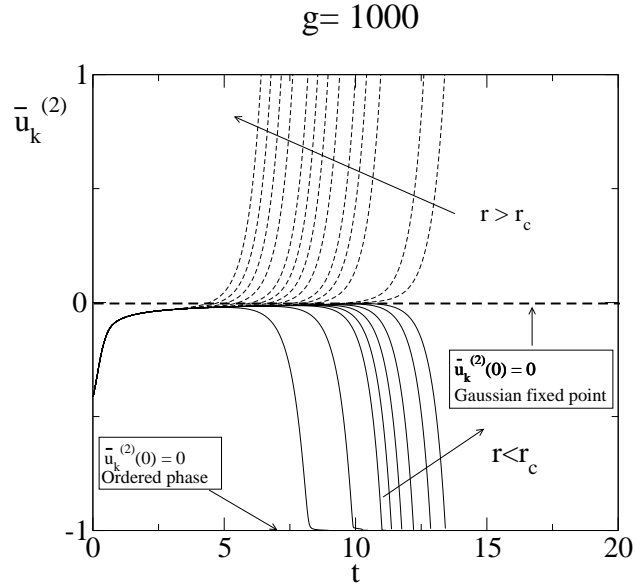}
\caption{
\label{u2G} 
The coupling constant $\overline{u}^{(2)}_k \equiv [ d^2 U_k(\phi=0)/d\phi^2] / \epsilon_k$
as a function of the RG time $t=\ln \Lambda /k$ at $g=1000$. For $r
>r_{c}$ the flow
escapes to infinity (dotted lines) while \mbox{, for $r
<r_{c}$ ,} the flow reaches the 
low temperature fixed point $\overline{u}^{(2)}_k = -1$ (solid lines). For $t \to \infty$
the dashed line $\overline{u}^{(2)}_k=0$ (Gaussian fixed point value) separates the two regimes.
}
\end{figure}

We pointed out  in section~\ref{Structure} that in the asymptotic limit $k \to 0$
the lattice and off-lattice LPA flow equations bear the same form.
In the ordered phase their behavior are both singular because of the simple pole $\omega=-1$ in 
the threshold function $\mathcal{L}(\omega)$ (see equation~\eqref{L}).  
This point has been studied at length in references~\cite{Bonanno,Caillol_I}.
Specializing this discussion to the case $D=4$ we note that in the limit $k \to 0$,
$\omega_k(\phi) = U_k^{''}(\phi)/ \epsilon_k \to -1$ for $-\phi_0(k)<\phi < \phi_0(k) $ where
$\phi_0(k)$ is a precursor of the spontaneous magnetization  $\phi_0 = \lim_{k \to 0}  \phi_0(k)$.
 It follows that the threshold function $\mathcal{L}$
diverge in this interval as $k^{-2}$. This yields a universal behavior $\mathcal{L}(\phi)/ \mathcal{L}(\phi=0)
= 1-\phi^2 /\phi_0^2$.  Moreover, as a consequence, $U_{k}(\phi)$ becomes convex as $k \to 0$, in particular
it becomes constant for $-\phi_0 < \phi < + \phi_0$.

The divergence of the threshold function makes impossible to obtain numerical solution of the non-linear 
PDE~\eqref{flow_LMD_final} in the ordered phase, we really deal with \textit{stiff} equations.
In order to remove stiffness, one is led to make the change of variables 
$U_k(M) \Longrightarrow L_k(M) = \mathcal{L}[   \omega_k(M) \equiv U_k^{''}(M)/\epsilon_k]$. We then obtain the equations
\begin{equation}
\label{flow_L}
L^{''}_k(\phi) =       \frac{2 \epsilon_k}{\mathcal{N}(\epsilon_k)}    \; 
\left[ \frac{1}{L_k(\phi) } -1 \right]     +
                                                                \frac{\epsilon_k}{\mathcal{N}(\epsilon_k)} \; \frac{1}{L_k(\phi)^2}
                                                                \; \partial_t L_k(\phi)
\end{equation}
where $k= \Lambda e^{-t}$.

In contradistinction with equations~\eqref{flow_LMD_final} the quasi-linear parabolic PDE~\eqref{flow_L} can easily be integrated out. 
As in references~\cite{Parola1,Bonanno,Caillol_I,Caillol_II}
we made use of the fully  implicit predictor-corrector algorithm of Douglas-Jones~\cite{Douglas}. This algorithm is unconditionally
stable and convergent and introduces an error of $\mathcal{O}( (\Delta t)^2 )  + \mathcal{O}((\Delta \phi)^2)$
($\Delta t$ and $\Delta \phi$ discrete RG time and field steps
respectively) and can be used below and above the critical point as well. 
In the ordered phase we note that~\cite{Caillol_I} $L_k(\phi) \propto k^{-2}(\phi_0(k)^2  - \phi^2) $ for
$-\phi_0 < \phi < + \phi_0$ which obviously does not precludes us to obtain a
numerical solution of equation~\eqref{flow_L}.

The initial conditions on the local potential $U_k$ at $k=\Lambda$ are easily
transposed to the field $L_k$. It follows from the discussion at the end of
section~\ref{loc} that the simplest choice is to choose 
$\Lambda=k_{\mathrm{max}}= 4 / a $ and  $L_{\Lambda}=\mathcal{L}(a^{-4} 
\gamma^{''}_{k_{\mathrm{max}}}(\overline{\phi})) $ where $\gamma_{k_{\mathrm{max}}}$ is the local
Wetterich function and $\overline{\phi}=a \phi$ for all values of the order parameter
$\phi$.

Of course, in practice, a cut-off must be imposed on $\phi$ and boundary conditions
must then be introduced such that the PDE is solved only
on the interval $-\phi_{\mathrm{max}} < \phi < \phi_{\mathrm{max}}$
for all $k$ with some specifications on the boundaries. We made
the consistent choice~\cite{Caillol_I,Caillol_II}
$L_k(\pm \phi_{\mathrm{max}}) =a^{-4} \mathcal{L}(a^{-4}\gamma{''}_{k}(\overline{\phi}_{\mathrm{max}}))$.
Here Wetterich effective function $\gamma_{k}(\overline{\phi}_{\mathrm{max}})$ is evaluated in the first
approximation of the hopping parameter expansion (see \textit{e.g.} reference~\cite{Montvay})
by assuming the validity of the local approximation.

\begin{figure}[!]
\centering
\includegraphics[angle=0,scale=0.40]{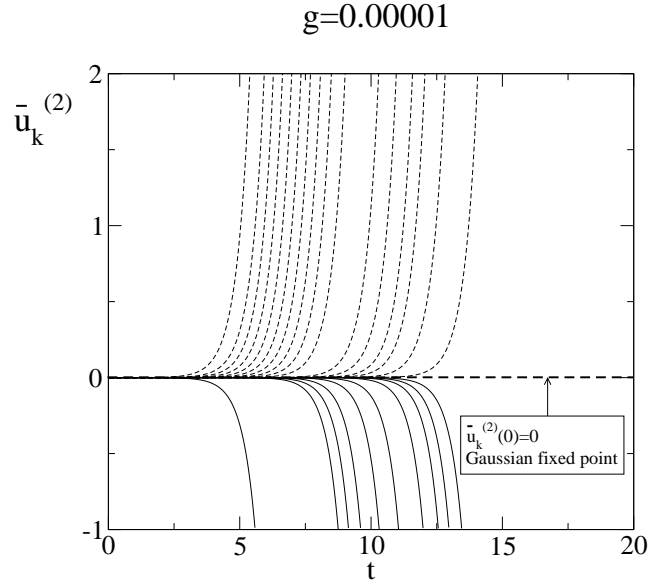}
\caption{
\label{u2g} Same as figure~\ref{u2G} for $g = 0.00001$
}
\end{figure}
\begin{table*}[!]
\centering
\caption{ \label{Tab}Critical parameters  of the $\Phi^4$ scalar field theory on a
$4D$ simple 
cubic lattice in the LPA approximation using 
the LMD regulator~\eqref{Litim}.
From left to right : $g$, $\overline{r}_{c}(g)$. 
The data were obtained by fixing $g$
and determining $\overline{r}_{c}(g)$ by dichotomy.
An uncertainty of at most $\pm 1$ affects the last digit.}
\begin{tabular}{ ||l | l || l | l || } 
\hline  
 $ g $    &      $ \overline{r}_{c}(g)  $          &     $g  $    & 
 $\overline{r}_{c}(g)  $   \\ \hline
 0.10 $10^{-4}$ & -0.7746694 $10^{-6}$ &  0.70 $10^{2}$  & -0.4200564 $10^{1}$  \\   \hline 
 0.10 $10^{-3}$ & -0.7746662 $10^{-5}$ &  0.75 $10^{2}$  & -0.4456839 $10^{1}$  \\ \hline 
 0.50 $10^{-3}$ & -0.3873318 $10^{-4}$ &  0.80  $10^{2}$ & -0.4709800 $10^{1}$  \\ \hline
 0.10 $10^{-2}$ & -0.7746600 $10^{-4}$ &  0.85  $10^{2}$ & -0.4959654 $10^{1}$  \\ \hline
 0.10 $10^{-1}$ & -0.7745977 $10^{-3}$ &  0.90  $10^{2}$ & -0.5206587 $10^{1}$  \\ \hline
 0.20 $10^{-1}$ & -0.1549056 $10^{-2}$ &  0.95 $10^{2}$  & -0.5450764 $10^{1}$  \\ \hline 
 0.30 $10^{-1}$ & -0.2323377 $10^{-2}$ &  0.100 $10^{3}$ & -0.5692335 $10^{1}$  \\ \hline
 0.40 $10^{-1}$ & -0.3097560 $10^{-2}$ &  0.110 $10^{3}$ & -0.6168189 $10^{1}$  \\ \hline
 0.50 $10^{-1}$ & -0.3871605 $10^{-2}$ &  0.120 $10^{3}$ & -0.6635096 $10^{1}$  \\ \hline
 0.60 $10^{-1}$ & -0.4645512 $10^{-2}$ &  0.130 $10^{3}$ & -0.7093852 $10^{1}$  \\ \hline
 0.70 $10^{-1}$ & -0.5419282 $10^{-2}$ &  0.140 $10^{3}$ & -0.7545135 $10^{1}$  \\ \hline
 0.80 $10^{-1}$ & -0.6192914 $10^{-2}$ &  0.150 $10^{3}$ & -0.7989528 $10^{1}$  \\ \hline
 0.90 $10^{-1}$ & -0.6966409 $10^{-2}$ &  0.160 $10^{3}$ & -0.8427538 $10^{1}$  \\ \hline
 0.10		& -0.7739766 $10^{-2}$ &  0.170 $10^{3}$ & -0.8859610 $10^{1}$ \\ \hline
 0.20		& -0.1546584 $10^{-1}$ &  0.180 $10^{3}$ & -0.9286136 $10^{1}$ \\ \hline
 0.30		& -0.2317839 $10^{-1}$ &  0.190 $10^{3}$ & -0.9707466 $10^{1}$  \\ \hline
 0.40		& -0.3087757 $10^{-1}$ &  0.200 $10^{3}$ & -0.1012391 $10^{2}$  \\ \hline
 0.50		& -0.3856355 $10^{-1}$ &  0.225 $10^{3}$ & -0.1114543 $10^{2}$  \\ \hline
 0.60		& -0.4623647 $10^{-1}$ &  0.250 $10^{3}$ & -0.1214183 $10^{2}$  \\ \hline
 0.70		& -0.5389649 $10^{-1}$ &  0.275 $10^{3}$ & -0.1311605 $10^{2}$  \\ \hline
 0.80		& -0.6154375 $10^{-1}$ &  0.300 $10^{3}$ & -0.1407051 $10^{2}$  \\ \hline 
 0.90		& -0.6917840 $10^{-1}$ &  0.350 $10^{3}$ & -0.1592776 $10^{2}$  \\ \hline
 0.10 $10^{1}$  & -0.7680056 $10^{-1}$  &  0.400 $10^{3}$ & -0.1772604 $10^{2}$  \\ \hline 
 0.15 $10^{1}$  & -0.1147289	       &  0.450 $10^{3}$ & -0.1947454 $10^{2}$  \\ \hline
 0.20 $10^{1}$  & -0.1523643	       &  0.500 $10^{3}$ & -0.2118027 $10^{2}$  \\ \hline
 0.25 $10^{1}$  & -0.1897212	       &  0.550 $10^{3}$ & -0.2284875 $10^{2}$  \\ \hline
 0.30 $10^{1}$  & -0.2268122	       &  0.600 $10^{3}$ & -0.2448441 $10^{2}$  \\ \hline
 0.40 $10^{1}$  & -0.3002422	       &  0.650 $10^{3}$ & -0.2609089 $10^{2}$  \\ \hline
 0.50 $10^{1}$  & -0.3727360	       &  0.700 $10^{3}$ & -0.2767124 $10^{2}$  \\ \hline
 0.60 $10^{1}$  & -0.4443624	       &  0.750 $10^{3}$ & -0.2922800 $10^{2}$  \\ \hline
 0.70 $10^{1}$  & -0.5151810	       &  0.800 $10^{3}$ & -0.3076338 $10^{2}$  \\ \hline
 0.80 $10^{1}$  & -0.5852432	       &  0.850 $10^{3}$ & -0.3227925 $10^{2}$  \\ \hline 
 0.90 $10^{1}$  & -0.6545945	       &  0.900 $10^{3}$ & -0.3377728 $10^{2}$  \\ \hline 
 1.00 $10^{1}$  & -0.7232751	       &  0.950 $10^{3}$ & -0.3525890 $10^{2}$  \\ \hline
 1.25 $10^{1}$  & -0.8922688            &  0.10  $10^{4}$ & -0.3672538 $10^{2}$  \\ \hline
 1.50 $10^{1}$  & -0.1057756 $10^{1}$  &   0.12  $10^{4}$ & -0.4246102 $10^{2}$  \\ \hline
 1.75 $10^{1}$  & -0.1220112 $10^{1}$  &   0.14  $10^{4}$ & -0.4802738 $10^{2}$  \\ \hline
 0.20 $10^{2}$  & -0.1379637 $10^{1}$  &   0.16  $10^{4}$ & -0.5346330 $10^{2}$  \\ \hline
 0.25 $10^{2}$  & -0.1691160 $10^{1}$  &   0.18  $10^{4}$ & -0.5879670 $10^{2}$  \\ \hline
 0.30 $10^{2}$  & -0.1993908 $10^{1}$  &   0.20  $10^{4}$ & -0.6404841 $10^{2}$  \\ \hline
 0.35 $10^{2}$  & -0.2289055 $10^{1}$  &   0.25  $10^{4}$ & -0.7691726 $10^{2}$  \\ \hline
 0.40 $10^{2}$  & -0.2577512 $10^{1}$  &   0.30  $10^{4}$ & -0.8953949 $10^{2}$  \\ \hline
 0.45 $10^{2}$  & -0.2860003 $10^{1}$  &   0.40  $10^{4}$ & -0.1144133 $10^{3}$  \\ \hline
 0.50 $10^{2}$  & -0.3137118 $10^{1}$  &   0.50  $10^{4}$ & -0.1391031 $10^{3}$  \\ \hline
 0.55 $10^{2}$  & -0.3409347 $10^{1}$  &   0.60  $10^{4}$ & -0.1637724 $10^{3}$  \\ \hline
 0.60 $10^{2}$  & -0.3677103 $10^{1}$  &   0.70  $10^{4}$ & -0.1884735 $10^{3}$  \\ \hline 
 0.65 $10^{2}$  & -0.3940740 $10^{1}$  &   0.10  $10^{5}$ & -0.2627898 $10^{3}$  \\ \hline
\end{tabular}
\end{table*}
\subsection{\label{Num_II}Solving the flow equations}
We solved equation~\eqref{flow_L} with the Douglas-Jones algorithm \cite{Douglas}. We
used   for most  our numerical experiments    
$\Delta t = 10^{-4}$, a maximum of $N_t=3\; 10^5$ time steps, $\Delta \phi = 10^{-4}$ and $N_{\phi}=30000$ field steps
(\textit{i. e.} $\phi_{\textrm{max}}=3.$).
Note that the functions $\mathcal{N}(\epsilon)$ and $\mathcal{D}(\epsilon)$ can be computed once for all with the desired
precision. 

In order to determine the critical point $r_c(g)$ one proceeds by dichotomy,
$g$ is fixed and one varies $r$. An illustration of the method is given in figure~\eqref{u2G} in the case $g=1000$.
The renormalized coupling constant 
$\overline{u}_k^{(2)} \equiv U_k^{''}(M=0)/ \epsilon_k  $, with $\epsilon_k=a^2 k^2$, 
discriminates the state of the system by its behavior in the limit $k \to 0$.

Of course the Gaussian fixed point, characterized by $\overline{u}_k^{(2)}=0$, is never reached
but approached only asymptotically for $r=r_c(g)$. As soon as $r \neq r_c(g)$ the flow
deviates from the fixed point due to the relevant fields. For $r < r_c(g)$ the coupling constant
$\overline{u}_k^{(2)} \to -1$ as $t$ increases; this is the expected behavior in the ordered phase.
For $r>r_c(g)$, $
\overline{u}_k^{(2)}\to +\infty$ when $k \to 0$ (and thus $\epsilon_k \to 0$)
since the compressibility $U_k^{''}(\phi)$ remains finite for all values of the 
order parameter $\phi$; the curves escape to $+\infty$ as can be seen on the figure.

A few dichotomies of $r$ thus yield a very precise estimate of $\overline{r}_c(g)$.
We checked that
our values for the parameters $\Delta t $,  $\Delta \phi$, etc give at least $8$ stable figures for 
$\overline{r}_{c}(g)$. 
We report only $7$ figures in the table~\ref{Tab} with the last figure rounded-up. Precision could be enhanced
with codes in quadruple precision, unfortunately no such public domain FORTRAN code exists for the calculation
of Bessel functions.
We explored a wide range of values of
parameters with $g$ varying in the range $g=10^{-5}$ (the Gaussian limit) up to $g=100000$ (Ising model limit),
see respectively figures~\ref{u2g} and~\ref{u2G}. 

\begin{figure}[!]
\centering
\includegraphics[angle=0,scale=0.40]{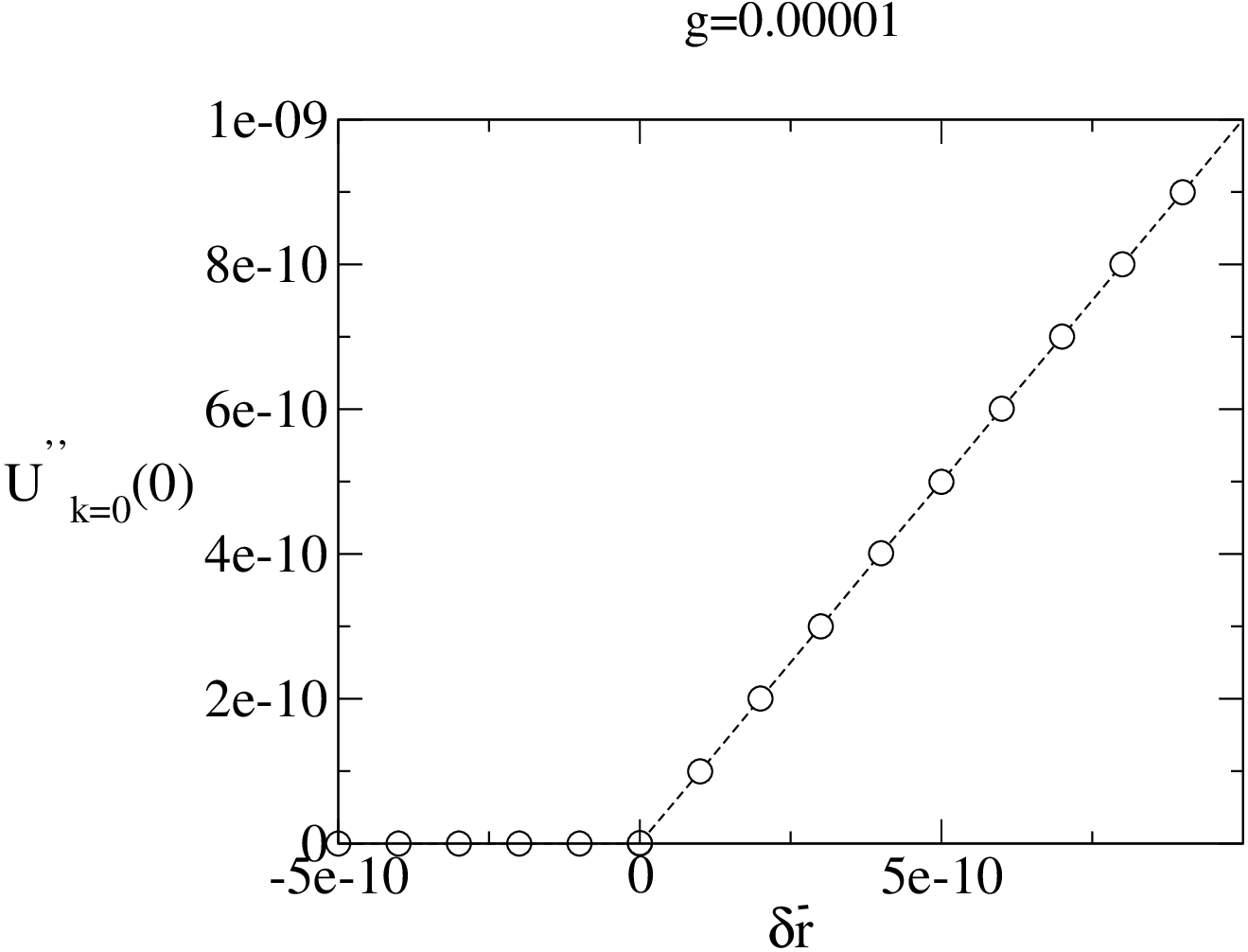}
\caption{
\label{Compr} Inverse compressibility $U^{''}_k(\phi=0)$ in the limit
$k \to 0$ for $g = 0.00001$ as a function of $\delta \overline{r} =
\overline{r} -\overline{r}_{c} $.
}
\end{figure}

Recent Monte Carlo simulations suggest, according to the authors of reference~\cite{Bordag}, 
the existence of a weak first order transition, at 
low values of $g$, \textit{i.e.} in the Gaussian limit.
Since there are no other fixed point (FP) than the Gaussian FP in $D=4$ it would mean that the flow stops
at some finite value of $k$ and does not reach the FP. Consequently hysteresis phenomena should
be observed in conjunction with the abortion of critical fluctuations.
This scenario is in contradiction with our findings in the LPA/LMPD theory. Figure~\ref{Compr} displays
the inverse compressibility $U^{''}_k(\phi=0)$ in the limit
$k \to 0$ for $g = 0.00001$. The fixed point is attained and the expected
linear classical behavior of $U^{''}_k(\phi=0) \Leftarrow (\delta \overline{r})$ is eventually obtained. 
A linear regression of the right part of the curve
gives an exponent of $\gamma^{-1} =0.99985$ in agreement with the classical value of the compressibility exponent
$\gamma=1$. A weak first order transition
would yield a discontinuity at some value of $r$ which is never observed for $g \geq 10^{-5}$. Numerically
it proved very difficult to consider smaller values of $g$ smaller than $10^{-5}$ and a code written in 
quadruple precision should be necessary to investigate further this question.
\section{\label{Conclusion} Conclusion}
In this paper we have computed the critical line of the $\Phi^4$ one-component model on the simple  cubic lattice in four dimensions
of space in the framework of the NPRG within the LPA approximation. We made use only the smooth LMD regulator which is expected
to give the better results.
The flow equations have been solved for the threshold functions rather than for the potential. This trick allows to obtain
numerical solutions in the ordered phase where the PDE for the potential are stiff and  fail to converge.  
A dichotomy process based on the generically different
asymptotic  behaviors of the dimensioned inverse susceptibility $U_k^{''}(\phi=0)/k^2$ in zero field, below and above the 
critical point,
yields a very precise determination of the  critical line $\overline{r}_c(g)$. The model is trivial in the sense that all the solutions
belong to the basin of attraction of the Gaussian fixed point for all considered values of $g$.
We did not observe a weak first order transition in the Gaussian limit
$g\to 0$, at least, numerically, for $g > 10^{-5}$. A numerical exploration of still lower values of parameter
$g$ would require  a quadruple precision code which is out of reach for the moment.

In reference~\cite{Caillol_II} we obtained an excellent agreement between our estimates of the critical line of the $3D$
$\Phi^4$ model  on a simple  three dimensional lattice
and that of Monte Carlo simulations of Hasenbush~\cite{Hasenbusch}. In $D=3$ the LPA approximation does not yield the
exact critical exponents contrary to the case $D=4$ where the classical exponents are found. One can thus \textit{a fortiori}
expect an excellent agreement for the critical line between the theory and the simulations in $4D$. Unfortunately, we were unable to find
estimates of the critical line of the $4D$ version of the model by means of Monte Carlo simulations in the literature.

\section*{Acknowledgments}
I warmly thank  Pr. D.~ Becirevic for interesting discussions and Pr. Y. L.~ Loh
for a fruitful exchange of e-mails. 


\lastpage
\end{document}